\documentclass{rnaastex}

\newcommand{\BL}{\textit{Breakthrough Listen}~}

\definecolor{black}{rgb}{0,0,0}
\definecolor{red}{rgb}{1.0,0,0}

\newcommand{\omm}{1I/$'$Oumuamua}
\begin{document}

\title{Breakthrough Listen Observations of \omm\ with the GBT}

\correspondingauthor{J. Emilio Enriquez}
\email{e.enriquez@berkeley.edu}


\author[0000-0003-2516-3546]{J. Emilio Enriquez}
\affiliation{Department of Astronomy, University of California, Berkeley}
\affiliation{Department of Astrophysics/IMAPP,Radboud University, Nijmegen, Netherlands}

\author[0000-0003-2828-7720]{Andrew Siemion}
\affiliation{Department of Astronomy, University of California, Berkeley}
\affiliation{Department of Astrophysics/IMAPP,Radboud University, Nijmegen, Netherlands}
\affiliation{SETI Institute, Mountain View, California}

\author{T. Joseph W. Lazio}
\affiliation{Jet Propulsion Laboratory, California Institute of Technology, Pasadena, California}

\author{Matt Lebofsky}
\affiliation{Department of Astronomy, University of California, Berkeley}

\author{David H.\ E.\ MacMahon}
\affiliation{Department of Astronomy, University of California, Berkeley}

\author{Ryan S. Park}
\affiliation{Jet Propulsion Laboratory, California Institute of Technology, Pasadena, California}

\author{Steve Croft}
\affiliation{Department of Astronomy, University of California, Berkeley}

\author{David DeBoer}
\affiliation{Department of Astronomy, University of California, Berkeley}

\author{Nectaria Gizani}
\affiliation{Department of Astronomy, University of California, Berkeley}
\affiliation{School of Science and Technology, Hellenic Open University, Greece}

\author{Vishal Gajjar}
\affiliation{Space Sciences Laboratory, University of California, Berkeley}

\author{Greg Hellbourg}
\affiliation{Department of Astronomy, University of California, Berkeley}

\author{Howard Isaacson}
\affiliation{Department of Astronomy, University of California, Berkeley}

\author[0000-0003-2783-1608]{Danny C.\ Price}
\affiliation{Department of Astronomy, University of California,  Berkeley}
\affiliation{Centre for Astrophysics \& Supercomputing, Swinburne University of Technology, Australia}



\section{Introduction.}
\label{sec:intro}

On 2017 October 18, the Pan-STARRS collaboration discovered an object within our solar system that appeared to be on a hyperbolic orbit \citep{Williams:2017tm}. Subsequent observations suggest this object (now designated 1I/2017 U1 or \omm) is of interstellar origin, lacks a coma, and that its shape is highly elongated relative to other known asteroids \citep{Meech2017}. 

It has long been suggested that advanced extraterrestrial civilizations, should they exist, could conceivably send probes to other stars either for exploration or communication purposes \citep{1960Natur.186..670B,1980JBIS...33...95F,2004Natur.431...47R,Gertz:2016wd}.  Interstellar probes would likely be equipped with communication technology that could potentially be operating in the radio band.

The \BL (BL) program \citep{2017AcAau.139...98W} conducted an observing campaign targeting \omm\ with the Robert C. Byrd Green Bank Telescope (GBT), with a goal of detecting, or placing limits on, radio emission consistent with a technological source. Because \omm\ is much closer to Earth than typical stellar targets, we have the opportunity to carry out a search for extremely weak transmitters: unprecedented in any other SETI experiment. Here we present a description of the observing campaign, and preliminary results.

\section{observations}
\label{sec:obs}

In December 2017, the BL Team conducted an initial 8-hour observation campaign using the GBT with four different receivers: L-band (1.1--1.9\,GHz), S-band (1.73--2.6\,GHz), C-band (4--8\,GHz) and X-band (8--11.6\,GHz).  Observations were conducted using the Breakthrough Listen back-end  \citep{MacMahonBL2017} to collect and analyze the data.

The 8-hour block was divided into two hours per receiver. Considering the suggested rotation of \omm, we subdivided the two hours per receiver into half-hour blocks, one for each phase quadrant of the rotation. Each half hour was further subdivided into six 5-minute observations in an ABACAD configuration as described in \cite{2017PASP..129e4501I} and \cite{2017ApJ...849..104E}. 

We set the time of the first observation as $t_{0}$  (2017 December 13 at 21:53:22 UTC; 58100.9121 MJD), and used a rotational period of 8.1 $\pm 0.02$ hrs from \cite{2017arXiv171104927B}
for the phase calculation. Over two weeks, we covered the full rotational phase of \omm\ with each receiver. Table \ref{table:obs_table} shows the MJD dates relating receiver band coverage to quadrant phase.

\begin{table*}
\centering
\caption{Observation MJD dates for a given quadrant/band combination. The quadrants Q1, Q2, Q3 and Q4 are defined in phase space as 0.-0.25, 0.25-0.5, 0.5-0.75, and 0.75-1.0 respectively.}

\begin{tabular}{lcccc}
\hline
& L-band  & S-band & C-band & X-band\\
\hline
\hline

\\
Q1 & 58110.0323 & 58100.9121 & 58106.0037 & 58100.9791 \\
\\
Q2 &58101.0313 
& 58106.0570 & 58107.1261 & 58117.9094\\
\\
Q3 & 58107.1768 & 58109.9187 & 58101.09403 & 58109.8753 \\
\\
Q4 & 58109.9994 & 58105.8901 & 58109.9749  & 58105.9473\\
\\

\hline
\end{tabular}
\label{table:obs_table}

\end{table*}

\section{Results and Discussion}
\label{sec:conclusion}

We conducted a search of the data for narrow-band (3Hz resolution) drifting sinusoids as described by \cite{2017ApJ...849..104E}, over a drift rate range of $\pm$2 Hz/s. This range includes any acceleration in the geocentric and barycentric frames, as well as accommodating a transmitter located anywhere on the body itself. Our preliminary results show no narrow band radio emission from the direction of \omm\ at any rotational phase.

The object was at $\sim 2$\,AU when observed, and given an approximate SEFD of 20 Jy, with a 300\,s observation, and a 5$\sigma$ threshold, these observations were sensitive to a hypothetical transmitter with an EIRP of $\sim 0.08$\,W \
($\sim$3,000 times weaker than \href{https://descanso.jpl.nasa.gov/DPSummary/090924dawn-FinalCorrex--update5G.pdf}{the Dawn spacecraft communication down-link}.).

Based on the possibility that \omm\ could be in fact a dormant comet with delayed outgassing, we also searched for any indication of hydroxyl emission at the four transitions between 1612\,MHz and 1720\,MHz. We searched the L-band data taken during quadrant Q2 by stacking the three 5-min observations. No emission was detected, confirming previous observations during closer approach that the nature of the object is consistent with an asteroid-like composition (Park 2017 in prep).\nocite{parkinprep}

All data collected by BL will be publicly available. A subset of the observations described here can be downloaded from the \href{http://breakthroughinitiatives.org/opendatasearch}{Breakthrough Listen Public Web Archive}.

\acknowledgments

Funding for BL is provided by the \href{http://breakthroughinitiatives.org}{Breakthrough Prize Foundation}. 
Part of this research was carried out at the Jet Propulsion Laboratory, California Institute of Technology, under a contract with the National Aeronautics and Space Administration.

\bibliographystyle{aasjournal}
\bibliography{references}

\end{document}